\documentclass{iau}
\usepackage{graphicx}

\title[Extragalactic Star Clusters] 
{Formation of extragalactic Star Clusters}

\author[G. Hensler, Y. Hein, A. Boselli \& the VESTIGE consortium]  
{Gerhard Hensler$^1$,
Yannick Hein$^1$,
Alessandro Boselli$^2$ \\
 \and the VESTIGE consortium
 \thanks{Based on observations obtained with MegaPrime/MegaCam, a joint project of CFHT and CEA/DAPNIA, at the Canada-French-Hawaii Telescope (CFHT) which is operated by the National Research Council (NRC) of Canada, the Institut National des Sciences de l'Univers of the Centre National de la Recherche Scientifique (CNRS) of France and the University of Hawaii}
 }

\affiliation{$^1$Department of Astrophysics, University of Vienna, Austria \\
email: {\tt gerhard.hensler@univie.ac.at} \\[\affilskip]
$^2$Laboratoire d'Astrophysique de Marseille \& Aix Marseille Univ, CNRS, CNES, France }

\pubyear{2026}
\volume{403}  
\jname{The Hidden Beauty of Galactic Outskirts}

\usepackage{graphicx}
\usepackage{txfonts}
\usepackage{url}
\usepackage{lscape}
\usepackage{longtable}
\usepackage{color}

\newcommand{\Msun}{M$_{\odot}$}
\newcommand{\SFR}{\Msun yr$^{-1}$}

\newcommand{\Ha}{H$_\alpha$ }
\newcommand{\HII}{H{\,\sc{ii}}}
\newcommand{\fesc}{f$_{esc}$ }

\begin{document}

\maketitle

\begin{abstract}

The outskirts of galaxies show their beauty by an assembly of Globular Clusters and satellite galaxies, but also by interactions with their environment. Galaxies in clusters experience dynamical effects by mutual encounters and by the intra-cluster gas. This latter exerts a drag to remove gas from the outskirts. These stripped clouds are capable to form stars and open fundamental questions on astrophysical processes, as e.g. the survival of intergalactic clouds, the star-formation process, the evolution of isolated star clusters, etc. Our studies target at star clusters around the Virgo cluster spiral galaxy NGC 4254, observed within the VESTIGE. 60 young clusters are identified in the GALEX/FUV with different masses, ages, and separations from the mature galactic disk. Only half of them are also emitting \Ha as long as the clusters are still encompassed by gas until ram pressure strips it off leading to star-formation quenching and LyC escape. 


\end{abstract}

\section{Introduction}

In the recent decades an increasing number of fainter extragalactic objects were detected from extragalactic HI clouds (ECGs) to faint star clusters (SCs), both in the field and in galaxy clusters. These latter are permeated by gas falling in from the cosmic web and shock-heated to virial temperatures of several 10$^7$ K. 
According to the cosmological accretion scenario, also galaxies are assembled in clusters and also move close to virial velocities of $\sim$ 1000 km/s. These extremely high velocities produce two main dynamical effects: close encounters of galaxies lead to tidal interactions and a drag force is exerted by the tenuous intra-cluster medium (ICM). While the first affects all galactic components gravitationally, the ram pressure unbinds solely gas from galaxies' outskirts inwards to radii at which the binding energy of the galactic gravity still exceeds the drag. 

HI and \Ha observations of galaxy disk in clusters prove these truncations (\cite[Morgan et al. 2004]{Mor24}) which are well described for face-on motions against the ICM by the ram-pressure stripping (RPS) study of \cite[Gunn \& Gott (1977)]{GG77}. In their simulations \cite[Roediger \& Hensler (2005)]{RH05} distinguish two main stages of stripping, at first, the full drag bends the gaseous outermost disk to the back and disrupts the stripped gas to smaller cloudlets of which some fall back in the wind shadow while most of them are decelerated to rest with the ICM. Secondly, when a galaxy on its orbit leaves the denser ICM, the drag decreases again and the remaining disk edge is affected by Kelvin-Helmholtz instability and escapes as gaseous filaments (\cite[Boselli et al. 2016]{Bos16}). Both, cloudlets and filaments shine at least in \Ha.

\section{Stability of extra-galactic clouds}

At first, the clouds survival must be considered. Since EGCs should be Dark Matter free, this cannot play a stabilizing role. Numerical models of the EGC-adequate High-velocity Clouds (HVCs) by \cite[Sander \& Hensler (2019)]{SH19}, however, demonstrate that EGCs even of relatively low masses and homogeneous initial conditions are stabilized by their enhancing  self-gravity, so that homogeneous HVC simulations in the literature neglecting the Poisson equation fail to represent realistic EGCs.
EGCs surrounded by hot ICM are exposed to thermal conduction (TC) and ionizing intergalactic radiation and should, thereby, become photo- or/and thermally evaporated. While investigations with classical Spitzer TC find that the mass flux can switch its direction from evaporation to condensation of surrounding hot gas for lower cloud temperatures and cloud radii much smaller than the Field length, this accretion is enhanced in the case of a more realistic saturated TC (\cite[Sander \& Hensler 2023]{SH23}).

Even with these results for EGCs as those found in isolation by \cite[Jozsa et al. (2022)]{Jozsa22}, the existence of 100 kpc HI tails like e.g. of NGC 4388 (\cite[Yagi et al. 2013]{Yagi13}), NGC 4254 (\cite[Boselli et al. 2018b]{Bos18b}), and the RPS dwarf galaxy ESO 137-001 (\cite[Sun et al. 2007]{Sun07}) is still puzzling. 

\vspace{-0.5cm}

\section{Star formation in ram-pressure stripped gas}

The fact that only a few cases of dark EGC are identified to harbor SCs rises fundamental questions in astrophysics: 
Are EGC originating from RPS stable? 
How do SCs form in EGCs? 
What is the SF timescale and efficiency?
How is the stellar initial mass function (IMF) sampled at low SFR? 
At which time doss the gas envelop decouple from the SC when it is further accelerated while the SC stays at the EGC velocity at its formation? 
If this happens not too late, the SC gets ''naked'' so that Lyman continuum (LyC) radiation can escape. 

Observations of galaxies in nearby galaxy clusters attest this RPS and in some detached gas clouds SF is in action (\cite[e.g. Giuchin et al. 2025, Gullieuszik et al. 2023]{Giu25,Gul23}) or witness the past event by truncated gas disks. Most prominently, as well studied proto-typical RPS galaxies, VCC~1217 (\cite[Fumagalli et al. 2011]{Fum11}) and ESO~137-001 (\cite[Waldron et al. 2023]{Wal23}) can be resolved by their distinct SF regions in the tails. Numerical simulations that allow for SF within the gas tails (\cite[e.g. Roediger et al. 2014]{Roe14}) overestimate the SFR and produce an excess of SCs. In a recent paper \cite[Steyrleithner et al. (2020)]{Ste20} study the structural dependence of RPS gas on the relative velocity and conclude that high speeds ($\sim$1000 km/s) can push out massive gas clumps to form stars in contrast to more filamentary tails at lower velocities (e.g. $\sim$200 km/s). 

\vspace{-0.5cm}

\section{Star formation in EGCs of NGC 4254}

\begin{figure}
   \centering
   \includegraphics[width=12cm]{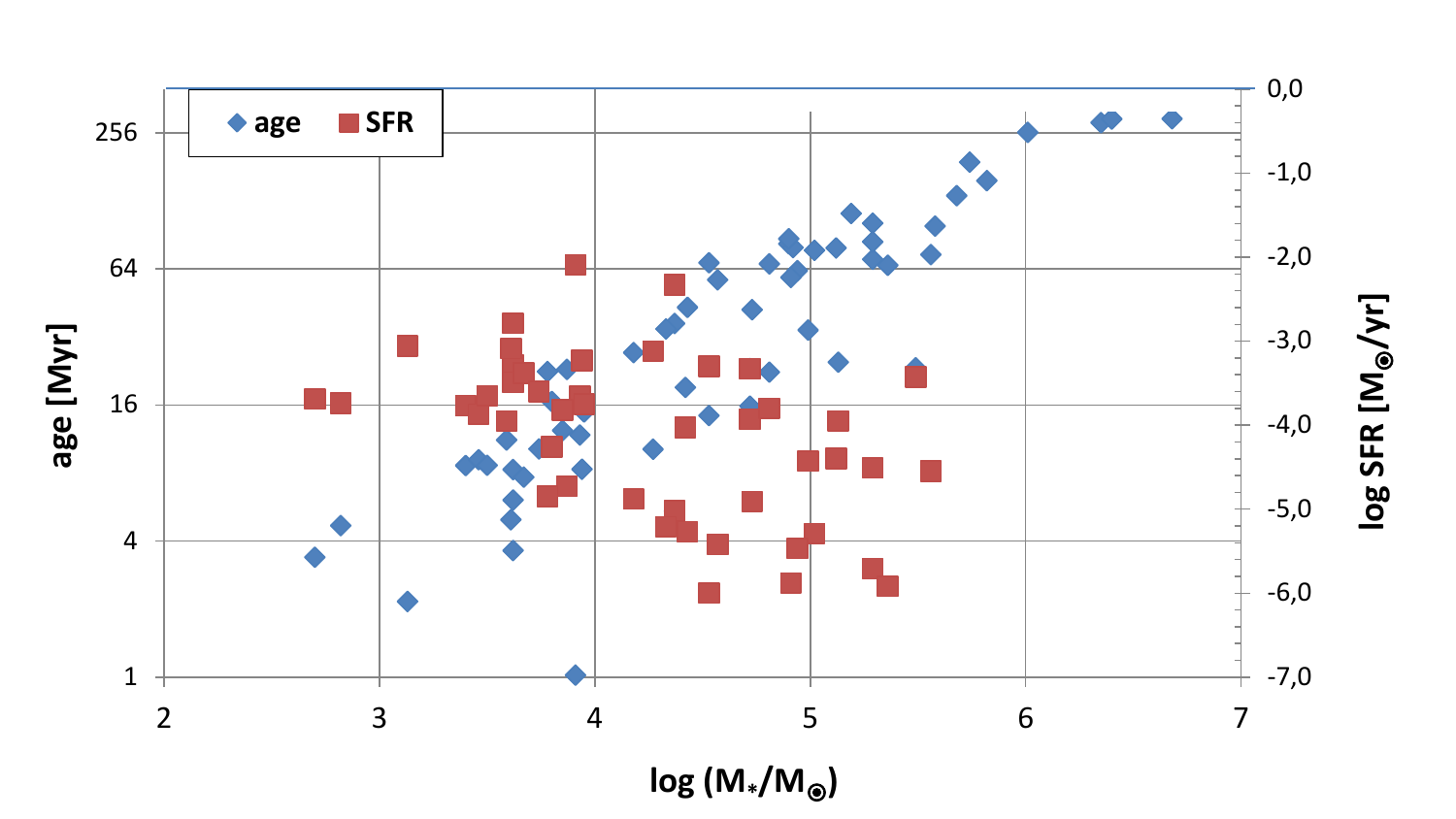}
   \caption{Ages and star-formation rates as function of the cluster masses. }
   \label{logage_SFR}
   \end{figure}
   
As an ideal target to address the above-mentioned problems, we detect the spiral (SA(s)c) galaxy NGC 4254 (M99) by VESTIGE (A Virgo Environmental Survey Tracing Ionised Gas Emission), a deep narrow-band \Ha imaging survey of the Virgo cluster carried out with MegaCam at CFHT (\cite[Boselli et al. 2018a]{Bos18a}). This survey is aimed at detecting diffuse ionized gas as e.g. stripped from perturbed galaxies and compact \HII\, regions of luminosity L(\Ha) $\geq$ 10$^{36}$ erg s$^{-1}$. 
The study is based on multi-spectral observations of NGC~4254 with VESTIGE and GALEX. Details of the VESTIGE survey and data reduction are given in \cite[Boselli et al. (2018a)]{Bos18a}. 

In combination with other SF tracers the \Ha line emission serves as a signature of a SF on timescales of $\sim$ 10-30 Myr. 
In addition, very deep GALEX exposures are available in the FUV band and in the NUV  (\cite[Boselli et al. 2011]{Bos11}) extending the age tracer of a simple stellar population (SSP) to $\sim$200 Myr. The specifications of the GALEX data are described in \cite[Boselli et al. (2018b)]{Bos18b}(hereafter:B18b). 

Since the SF in the stripped EGCs must happen almost instantaneously and is quenched due to the gas removal, a short episode of
${SFR(t)=t\times e^{-\frac{t}{\tau_{sf}}}}$\SFR
is applied with $\tau_{sf}$=3 Myr, 

The SC ages have been derived in B18b using the CIGALE SED fitting code for the population synthesis model (\cite[Bruzual \& Charlot 2003]{BC03}) with a Salpeter IMF. There is no distinct dependence of SC masses and ages with the separation from the galaxy (Hensler et al. 2026). Most ages are in the range below 20 Myrs with an extension to 100 Myrs and only 4 older than 250 Myrs (fig. \ref{logage_SFR}). The masses concentrate to $10^4-10^5$ \Msun, while the 4 oldest are the most massive with log(M/\Msun)$>$6. It becomes also obvious in fig. \ref{logage_SFR} that a mass-age dependence exists in the sense that the more massive clusters are older. This fact is e.g. also derived for the SCs in the tail of ESO~137-001 by \cite[Waldron et al. (2023)]{Wal23} and reasons are discussed in Hensler et al. (2026).

The SFRs are very low, mostly between $10^{-4}-10^{-3}$\SFR and even lower. This means that the mass distribution within the IMF must be compiled by random sampling or by an uppermost mass truncation, so that a full IMF must be questioned.   

The SC ages are applied to derive stellar fluxes as a function of stellar mass. Ionizing photon fluxes are taken from \cite[Sternberg et al. (2003)]{Ste03}, converted to \Ha luminosities using \cite[Calzetti et al. (2013)]{Cal13} and subsequently back to the LyC by means of the pseudo-filter introduced by \cite{Bos11}. UV flux data were provided by Luciana Bianchi (private comm.). The individual mass-dependent stellar fluxes are summed-up to absolute SC fluxes and scaled to the distance of the Virgo cluster (as M87 at 16.7 Mpc). The stellar masses are distributed over time by numerically integrating the SFR and populating 80 mass bins. 1000 IMF sampling iterations are performed for each cluster to ensure statistical robustness and will be published in Hein et al. (in prep.).

The correlation that the more massive SCs are older, allows the conclusion that the most massive cloud complexes are pushed off by the ram pressure at first, while the smaller clouds follow. The physical reason for this fact remains to be explored. For the most massive SCs the SFR is too uncertain and dropped in fig. \ref{logage_SFR}.

\vspace{-0.5cm}

\section{Timescale of gas removal}

\begin{figure}
   \centering
   \includegraphics[width=12cm]{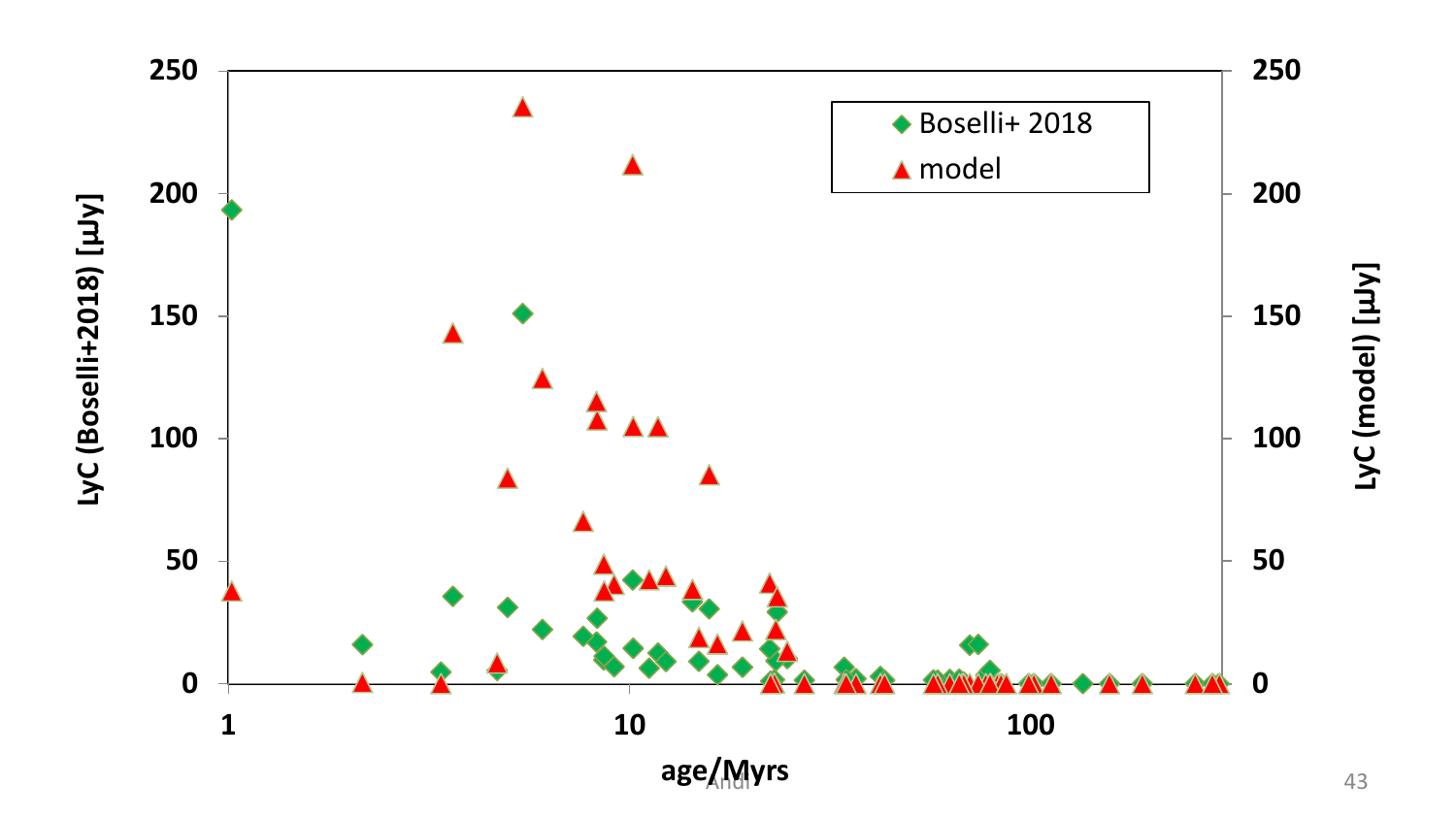}
   \caption{Comparison of the LyC flux as derived in (Bos18b) with the model of a SSP. The LyC(model) fluxes drop to zero because they vanish with the aging of a SSP.}
    \label{LyC}
\end{figure}

If one extrapolates the conditions of high-velocity cloud simulations by \cite[Sander \& Hensler (2021)]{SH21} or applies the ICM conditions to eq. (15) in \cite[Mori \& Burkert (2000)]{MB00}, one yields a gas removal timescales of 2-5 Myrs. Additionally, the internal SC energetics provided by supernovae typeII (SNeII) would expel the gas and quench the SF within a short time of atleast 3 Myr, but if the IMF lacks of most massive and shortest-living stars, what is reasonable for very low SFRs, this SNII pulse would be delayed and the RPS does the job.

For EGCs the exhaustion of the surrounding gas can be derived from the vanishing \Ha emission, when the SC is still young and should release LyC photons. From fig.\ref{LyC} the drop of the observed LyC emission (green rhombs) happens before 10 Myr. Only until ~3 Myr the observationally derived LyC fluxes exceed the model ones,  while the excessive model values (red triangles above green rhombs) reflect the lost of the gaseous envelops. The LyC escape fraction \fesc can then be derived by \fesc = 1-[LyC(Bos18)-LyC(model)] and varies from zero to 80\% depending on age, EGC mass, and upper-mass IMF (see \cite[Hensler et al. 2026]{Hen26}). While the model LyC emission ceases when all stars more massive than $\sim$ 15 \Msun have died (red triangles at zero in fig. \ref{LyC}), the measurable LyC (green rhombs) for older EGCs can only be interpreted by the assumption that $\tau_{sf}$ is longer than 3 Myr and more extended.

\end{document}